\documentclass[12pt,preprint]{aastex}
\newcommand{\hs}{\mbox{HS\,2325}}
\newcommand{\porb}{\mbox{$P_{\mathrm{orb}}$}}
\newcommand{\msun}{\mbox{$\mathrm{M}_{\odot}$}}
\newcommand{\rsun}{\mbox{$\mathrm{R}_{\odot}$}}
\newcommand{\mwd}{\mbox{$M_{\mathrm{WD}}$}}

\newcommand{\msec}{\mbox{$M_{\mathrm{sec}}$}}
\newcommand{\kwd}{\mbox{$K_{\mathrm{WD}}$}}
\newcommand{\ksec}{\mbox{$K_{\mathrm{sec}}$}}
\newcommand{\kms}{\mbox{$\mathrm{km\,s^{-1}}$}}

\newcommand{\Ion}[2]{#1{\,\scriptsize #2}}

\shorttitle{HS\,2325+8205}
\shortauthors{Pyrzas et al.}

\begin{document}

\title{HS\,2325+8205 - an ideal laboratory for accretion disc physics}

\author{S. Pyrzas\altaffilmark{1}, B. T. G\"ansicke\altaffilmark{1}, J. R. Thorstensen\altaffilmark{2}, A. Aungwerojwit\altaffilmark{3,4}, 
D. Boyd\altaffilmark{5}, S. Brady\altaffilmark{6}, \\ J. Casares\altaffilmark{7,8}, R. D. G. Hickman\altaffilmark{1}, T. R. Marsh\altaffilmark{1},
I. Miller\altaffilmark{9}, Y. \"Ogmen\altaffilmark{10}, J. Pietz\altaffilmark{11}, \\ G. Poyner\altaffilmark{12}, P. Rodr\'iguez-Gil\altaffilmark{7,8},
B. Staels\altaffilmark{13}} 

\altaffiltext{1}{Department of Physics, University of Warwick, Coventry, CV4 7AL, UK}
\altaffiltext{2}{Department of Physics and Astronomy, 6127 Wilder Laboratory, Dartmouth College, Hanover, NH 03755, USA}
\altaffiltext{3}{Department of Physics, Faculty of Science, Naresuan University, Phitsanulok, 65000, Thailand}
\altaffiltext{4}{ThEP Center, CHE, 328 Si Ayutthaya Road, Bangkok, 10400, Thailand}
\altaffiltext{5}{British Astronomical Association, Variable Star Section, West Challow OX12 9TX, UK}
\altaffiltext{6}{AAVSO, 25 Birch Street, Cambridge, MA 02138, USA}
\altaffiltext{7}{Instituto de Astrof{\'i}sica de Canarias, V{\'i}a L{\'a}ctea, s/n, La Laguna, E-38205, Tenerife,  Spain}
\altaffiltext{8}{Departamento de Astrof{\'i}sica, Universidad de La Laguna, E-38206 La Laguna, Tenerife, Spain}
\altaffiltext{9}{British Astronomical Association, Variable Star Section, Furzehill House, Ilston, Swansea SA2 7LE, UK}
\altaffiltext{10}{Green Island Observatory, Ge\c{c}itkale, Mugosa, via Mersin 10, Cyprus}
\altaffiltext{11}{Nollenweg 6, 65510 Idstein, Germany}
\altaffiltext{12}{British Astronomical Association, Variable Star Section, 67 Ellerton Road, Kingstanding, Birmingham B44 0QE, UK}
\altaffiltext{13}{CBA Flanders, Alan Guth Observatory, Koningshofbaan 51, Hofstade, Aalst, Belgium}


\begin{abstract}
We identify HS\,2325+8205 as an eclipsing, frequently outbursting dwarf nova with an orbital period of $\porb\,=\,279.841731(5)\,\mathrm{min}$. Spectroscopic 
observations are used to derive the radial velocity curve of the secondary star from absorption features and also from the H$\alpha$ emission lines, 
originating from the accretion disc, yielding $\ksec\,=\,K_{\mathrm{abs}}\,=\,237\,\pm\,28\,\kms$ and $K_{\mathrm{em}}\,=\,145\,\pm\,9\,\kms$ respectively. 
The distance to the system is calculated to be $400 (+200, -140)$ pc. A photometric monitoring campaign reveals an outburst recurrence time of 
$\sim\,12-14\,\mathrm{d}$. The combination of magnitude range ($17\,-\,14$ mag), high declination, eclipsing nature and frequency of outbursts makes 
HS\,2325+8205 the ideal system for ``real-time'' studies of the accretion disc evolution and behaviour in dwarf nova outbursts.
\end{abstract}

\keywords{stars}


\section{Introduction}

Dwarf novae are a sub-class of non- (or weakly) magnetic cataclysmic variables \citep[CVs, see e.g.][for a comprehensive review]{warner95-1}, in which a white 
dwarf primary accretes matter via an accretion disc, formed by material transferred through the $L_1$ point from a Roche lobe-filling (near) main-sequence 
secondary. The defining trait of dwarf novae are quasi-periodical brightness changes of several magnitudes, commonly known as ``dwarf nova outbursts''. It is 
widely accepted that outbursts can be understood within the framework of the disc instability model \citep[DIM, see e.g.][for reviews of the topic]{smak84-1,cannizzo93-1,osaki96-1,lasota01-1}. 
Within DIM accretion discs undergo outbursts if the mass transfer rate is below a critical value, $\dot{M}_{\mathrm{crit}}$. Above the CV orbital period 
gap\footnote{The orbital period range $2\,\rm{h}\,\leq\,\porb\,\leq\,3\,\rm{h}$, where only a small number of CVs are found.} accretion rates are usually 
larger than $\dot{M}_{\mathrm{crit}}$ and, as a result, only about a third of non-magnetic systems are dwarf novae. The situation is completely different 
below the period gap, where dwarf novae dominate the CV population \citep{shafter92-1}.

Dwarf novae provide the best environment to develop and test our understanding of accretion disc structure and dynamics, which is relevant to a wide range of
objects, such as low-mass X-ray binaries \citep[LMXBs,][]{dubusetal01-1}, active galactic nuclei \citep[AGN,][]{burderietal98-1} and young stellar objects 
\citep[YSO,][]{belllin94-1}.

Of particular interest in this context are eclipsing dwarf novae. In these systems, the physical properties of the binary, such as the mass ratio, the 
inclination angle, the masses and temperatures of the component stars and the radial structure of the accretion disc can be determined to high precision, 
through studies of the eclipse features of the white dwarf, the bright spot (formed in the region where the mass-transferring stream meets the accretion disc) 
and accretion disc components \citep[see e.g.][]{woodetal89-1,littlefairetal06-2,southworthetal09-1}.

HS\,2325+8205 (RA: 23$^{\mathrm{h}}$ 26$^{\mathrm{min}}$ 50.4$^{\mathrm{s}}$, Dec: +82$^{\circ}$ 22' 12'' [J2000], henceforth \hs) was one of the systems 
identified in a dedicated search for CVs \citep{aungwerojwitetal05-1} within the Hamburg Quasar Survey \citep[HQS,][]{hagenetal95-1}). Photometric observations 
soon revealed the eclipsing nature of the system and also frequently occuring outbursts. An interesting historic note is that \citet{morgentroth36-1} mentioned 
short-term variability of \hs, which correspondingly was included in the New Catalogue of Suspected Variable Stars as NSV\,14581 (though with rather uncertain 
coordinates).


\section{Observations}
\label{sec:obsall}

We obtained photometric and spectroscopic data on \hs\, using both large aperture ($>\,1\rm{m}$) and small aperture telescopes. Table\,\ref{tab:jobs} 
summarizes the observations conducted with the former. A brief account on data reduction follows.

\subsection{Photometry}
\label{subsec:obsphot}

\begin{figure}
\plotone{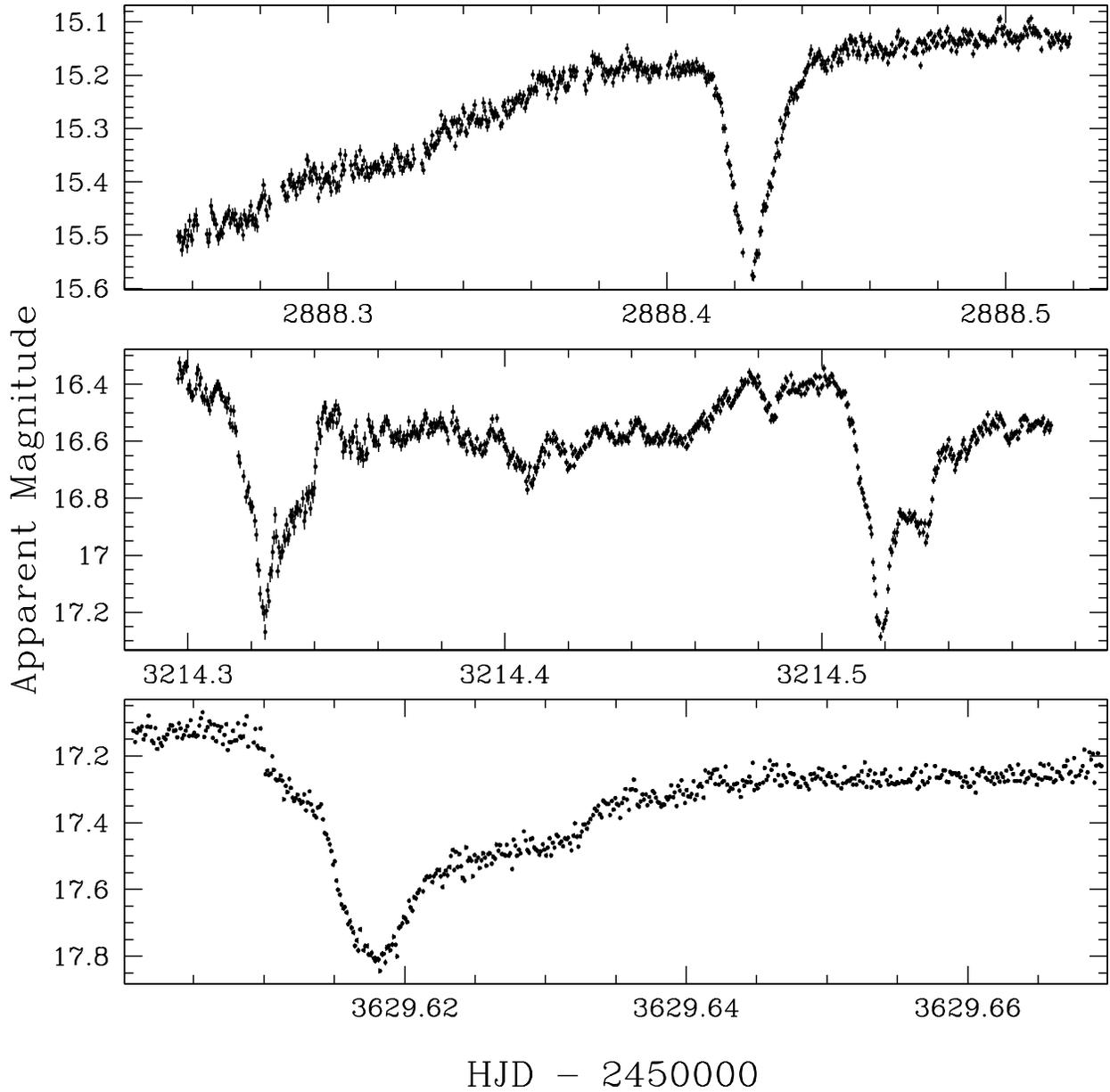}
\caption{Sample light curves of \hs. Top panel: filterless KY observations, from September 5, 2003 (ID01), with the system on the rise to outburst.  Middle 
panel: filterless KY observations, from July 27, 2004 (ID06), with the system in an intermediate state, Bottom panel: filterless NOT observations, from 
September 15, 2005 (ID13), with the system in quiescence. \label{fig:slc}}
\end{figure}

We obtained time-series photometry of \hs\ during 17 nights throughout the period 2003 to 2007 using 1.2--2.5\,m telescopes (Table\,\ref{tab:jobs}). These 
observations were reduced with the pipeline described in \citet{gaensickeetal04-1}, which employs bias-subtraction and flat-fielding in the standard fashion
within {\sc midas} and uses {\sc sextractor} \citep{bertin+arnouts96-1} to perform aperture photometry. Sample light curves are shown in Fig.\,\ref{fig:slc}. 

\hs\, has been found to vary in brightness between $\sim\,17^{th}$ and $\sim\,14^{th}$ magnitude. Eclipses are shallow and maintain an almost constant depth, 
during the rise to outburst. The eclipses in the bright state exhibit a symmetric U-shape, typical for an accretion disc-dominated system. During quiescence 
the eclipse morphology becomes more complicated and reveals several breaks in slope. In addition to eclipses, the light curve of \hs\ displays two further 
features: short-term, random, out-of-eclipse variations, known as ``flickering'' \citep[e.g.][]{bruch00-1}, and an ``orbital hump'', a brightening just before 
the start of the eclipse attributed to the bright spot coming into view \citep[e.g.][]{krzeminski65-1}.

\begin{figure}
\plotone{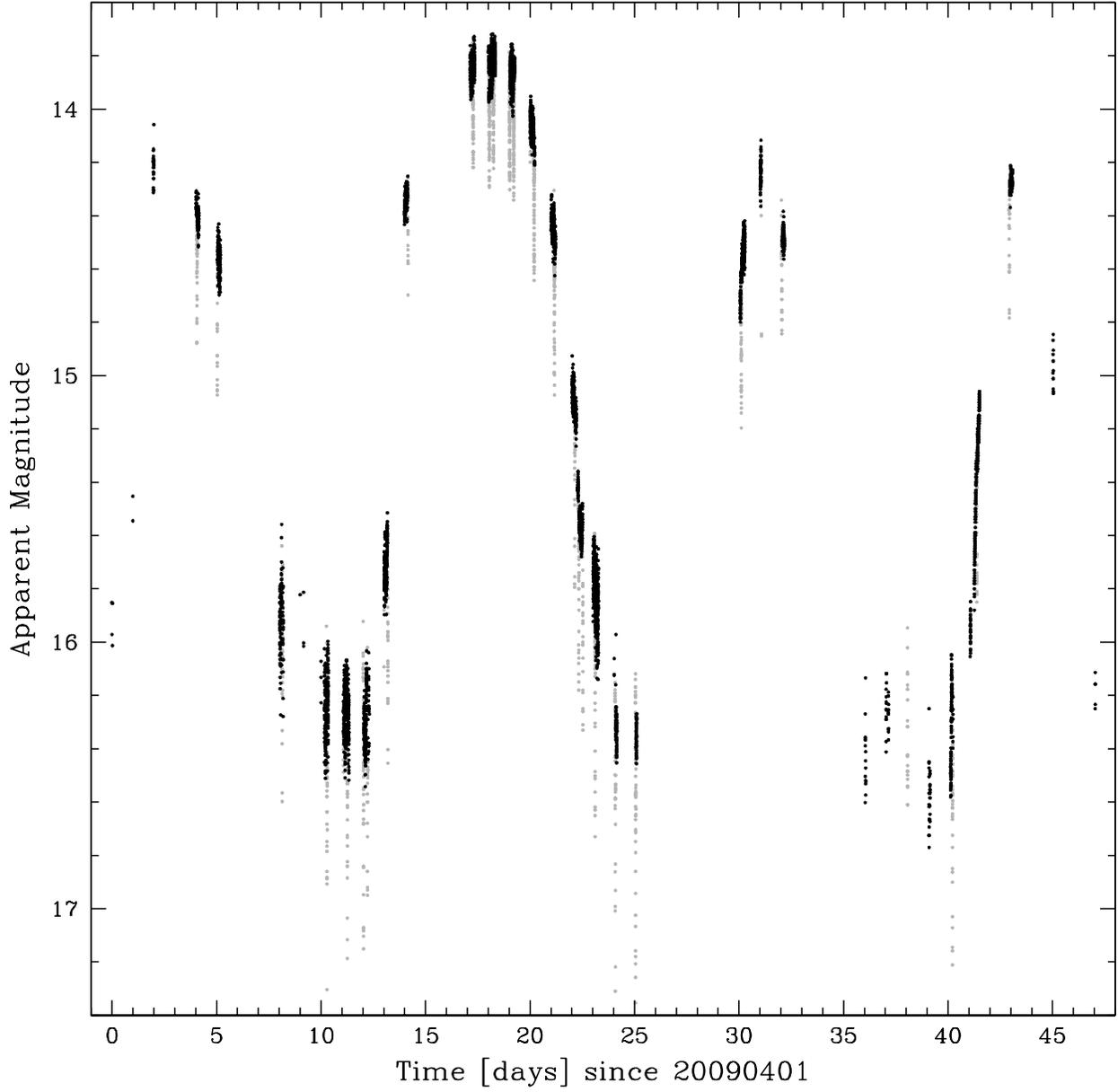}
\caption{The result of the outburst monitoring campaign for \hs. Four outbursts have been recorded in 50 days. Points in light grey indicate the system 
being in eclipse. \label{fig:outburst}}
\end{figure}

An intensive 1.5-month-long photometric campaign was conducted in 2009 to characterise the outburst behaviour of \hs, using small aperture (11"-14") telescopes. 
The data were reduced with {\sc aip4win} and {\sc maximdl}, and the resulting light curve is shown in Fig.\,\ref{fig:outburst}.

\subsection{Spectroscopy}
\label{subsec:obsspec}

\begin{figure}
\plotone{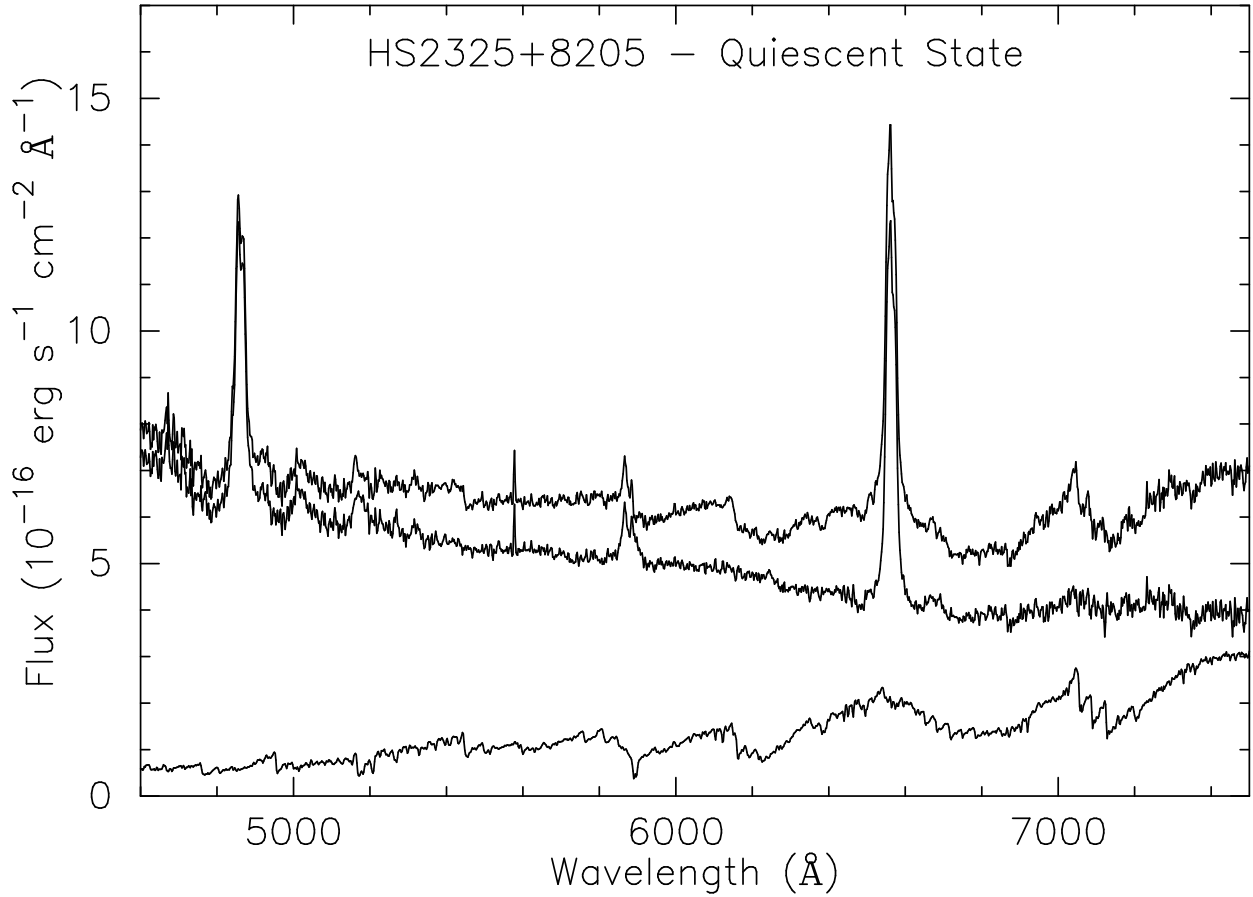}
\caption{Top trace: Mean quiescent spectrum of \hs. Bottom trace: A library spectrum of the M3 dwarf Gliese 436, scaled so that it has an apparent $V$-band 
magnitude of $\sim\,19.0$. Middle trace: The \hs\, spectrum minus the scaled M-dwarf. \label{fig:decomp}}
\end{figure}

Spectroscopic observations during the system's quiescence were obtained at the 2.4m Hiltner telescope at MDM Observatory on Kitt Peak, Arizona. The modular 
spectrograph and a SiTe 2048$^2$ pixel CCD yielded 2\,\AA/pixel and from 4210 to 7500\,\AA\ but with decreased sensitivity toward the ends of the wavelength 
range. The spectral resolution was $\sim\,3.5$\,\AA\ full width at half maximum (FWHM). Reductions were performed mostly with standard \texttt{IRAF} routines, 
but we used an original implementation of the optimal extraction algorithm detailed by \citet{horne86-1} to compute one-dimensional spectra from the 
two-dimensional images.  For wavelength calibration, we used a dispersion curve derived from lamp exposures in twilight, and corrected for nighttime drifts 
using the $\lambda\,5577$ sky line. We observed standard stars in twilight whenever the sky appeared clear, and used these observations to flux-calibrate the 
data. The scatter of the standard stars typically suggests that the flux calibration is uncertain by several tenths of a magnitude, probably due to 
uncalibrated losses at the spectrograph slit.  The mean quiescent spectrum is shown in Fig.\,\ref{fig:decomp}. The flux level of the observed spectrum implies 
a $V$-band magnitude near 17.0, subject to the calibration uncertainties.


\section{Orbital Period and Ephemeris}
\label{sec:porbneph}

Mid-eclipse times (given in Table\,\ref{tab:mideclt}) were determined by visually cross-correlating each eclipse profile with its mirror image with respect to 
time. This was found to produce more robust results than fitting a parabola to the eclipse minimum, in particular for the light curves with poor time 
resolution. We adopted the duty cycle (exposure plus readout time) of the corresponding observations as a conservative estimate of the uncertainty in the 
mid-eclipse times.

Fitting a linear ephemeris to the mid-eclipse times gives

\begin{equation}
\label{eq:ephe}
 T_{\rm{0}}(\rm{HJD})\,=\,2\,452\,888.42554(3)\,+\,0.194\,334\,535(3)\,E
\end{equation}

\noindent
with mid-eclipse times calculated on a UTC timescale, i.e. an orbital
period of $\porb\,=\,279.841731(5)\,\mathrm{min}$.


\section{Secondary spectral type and radial velocities analysis}
\label{sec:rvsec}

As is typical for quiescent dwarf novae, the Balmer lines in emission are the most prominent features in the spectrum of \hs\, with equivalent widths of 
$\simeq30$ and $\simeq54$\,\AA\ for H$\beta$ and H$\alpha$ respectively. \Ion{He}{I} emission is detected at 4921, 5015, and 5876 \AA , and \Ion{Fe}{II} 
$\lambda 5169$ as well (the features at $\lambda\lambda$ 4921 and 5015 may also be blended with \Ion{Fe}{II}). The absorption bands of an M-dwarf 
companion are conspicuous. To quantify the M dwarf contribution, we subtracted library spectra of M 
dwarfs classified by \citet{boeshaar76}, taken with the same instrument, and varied the spectral type and scaling until the M-dwarf features were cancelled as 
well as possible.  The lower two traces in Fig.\,\ref{fig:decomp} show the decomposition that was (at least subjectively) the best. From this exercise, we 
estimate that the companion is of type $\mathrm{M}3.0\,\pm\,0.75$ subclasses, and that its flux corresponds to $V\,=\,19.0\,\pm\,0.4$ (external error, 
including calibration uncertainties).  The spectral type-period relation of \citet{smith+dhillon98-1} (Equation\,4 in their paper) for 
$\rm{P}\,>\,4\,\rm{h}$ yields Sp2\,=\,M1.5 for the derived orbital period of $\porb\,=\,4.664\,\rm{h}$, a value broadly consistent with our estimate of
$\mathrm{M}3.0\,\pm\,0.75$, as the rms scatter of the spectral type-period relation is 3 subtypes for $\rm{P}\,>\,4\,\rm{h}$ \citep{smith+dhillon98-1}.

We measured radial velocities of the H$\alpha$ emission line using a double-Gaussian convolution method outlined by \citet{schneider+young80-2}; the centres of 
the Gaussians were separated by $1280\,\kms$, and each individual Gaussian had a FWHM of $270\,\kms$, comparable to our spectral resolution. This emphasised 
the outer wings of the line profile.  We also tried a range of separations, and found that the radial velocity amplitude and phase were insensitive
to this parameter.  To measure the velocity of the M-dwarf component, we used the cross-correlation program \texttt{rvsao}, written by 
\citet{kurtz+mink98-1}. For the template, we used a velocity-compensated composite M-dwarf spectrum, composed by summing the spectra of a large number of M 
dwarfs for which \citet{marcyetal87-1} tabulate precise velocities. The cross-correlation region was from 6000 to 6500 \AA ; this was chosen to include some 
strong atomic and TiO features, while avoiding emission lines. Not all the spectra gave usable cross-correlation velocities; we limited our analysis to those 
for which the formal velocity error was less than $35\,\kms$. 

\begin{figure}
\plotone{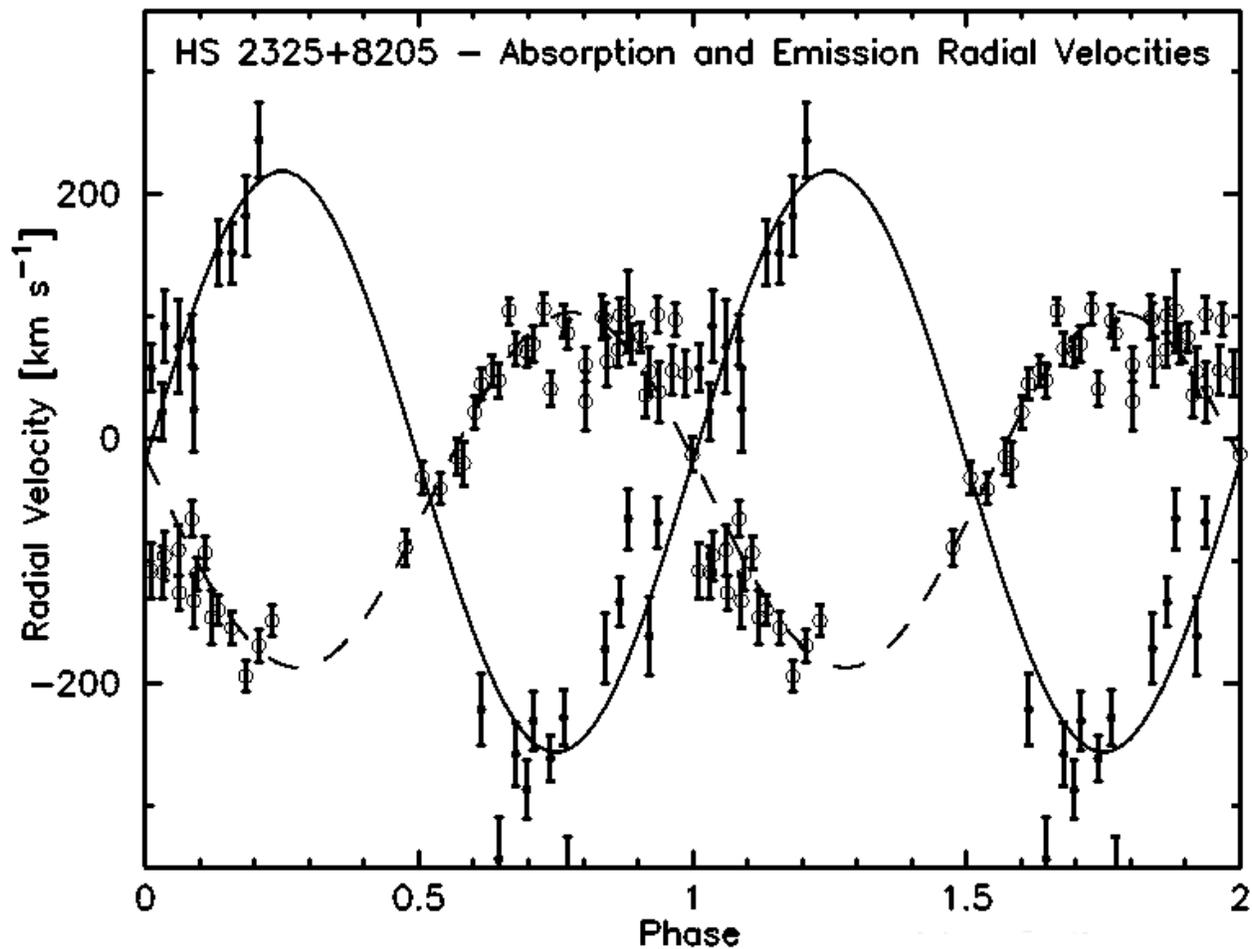}
\caption{Radial velocities of \hs\, in the quiescent state, plotted as a function of orbital phase. The open circles show the H$\alpha$ emission and their 
formal uncertainties, and the solid dots show the cross-correlation velocities of the M dwarf. There is a gap in coverage at $0.25\,<\,\phi\,<\,0.42$, and 
only some of the spectra yielded usable absorption velocities. The dashed curve shows the best fit to the emission velocities, with \porb\, fixed but 
$T_0$, $K$, and $\gamma$ allowed to vary. For the absorption velocity fit (solid line), the $T_0$ was held fixed to the eclipse phase, $K$ and $\gamma$ were 
adjusted. \label{fig:folpl}}
\end{figure}

\begin{figure}
\plotone{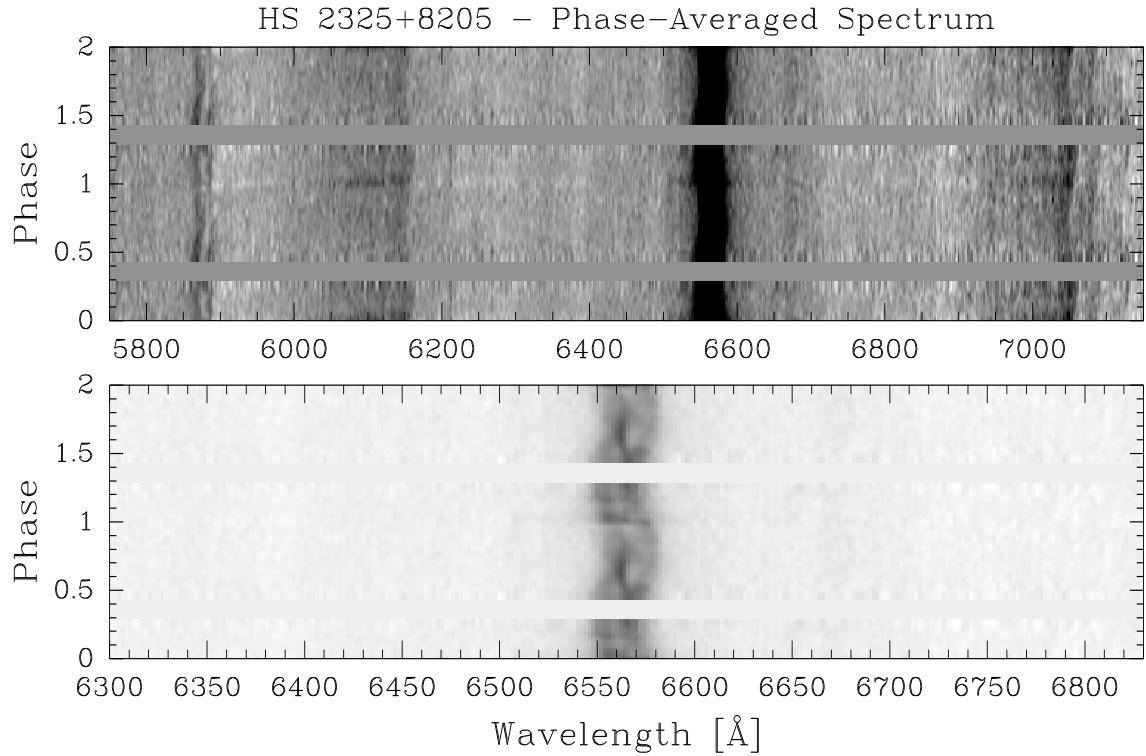}
\caption{Greyscale representation of the low-state spectra, presented as a function of orbital phase. The data are repeated for a cycle to preserve continuity. 
The two panels show the same data, but differ in the choice of greyscale limits. The horizontal blank bars are holes in the phase coverage. To create the 
figure the spectra were first rectified and cleaned of remaining cosmic ray hits; each line of the figure is a weighted average of spectra nearby in phase to 
the line's nominal phase, the weights being computed using a narrow gaussian. \label{fig:singletrail}}
\end{figure}

We then performed fits to the radial velocities (both absorption and H$\alpha$ emission), of the form $v(t)\,=\,\gamma\,+\,K\,\sin[(t\,-\,T_0)\,/\,\porb].$ The 
orbital period \porb\, was held fixed to the value derived from eclipses. Because of the modest number of absorption velocities and their limited phase 
coverage, and because the absorption should trace the motion of the secondary star fairly well, we fixed $T_0$ to the mid-eclipse ephemeris when fitting the 
absorption velocities, but left it as a free parameter for the H$\alpha$ emission ones. The resulting velocities were 
$\ksec\,=\,K_{\mathrm{abs}}\,=\,237(28)\,\kms$, $\gamma_{\mathrm{abs}}\,=\,-19(20)\,\kms$,  
$K_{\mathrm{em}}\,=\,145(9)\,\kms$, $\gamma_{\mathrm{em}}\,=\,-42(6)\,\kms$ for the absorption and emission lines respectively, with the numbers in parentheses 
indicating the errors. Figure\,\ref{fig:folpl} shows the emission and absorption velocities as a function of orbital phase, while Fig.\,\ref{fig:singletrail} 
shows a greyscale representation of the low-state spectra, as a function of phase. The upper panel of Fig.\,\ref{fig:singletrail} is scaled to emphasize the 
M-dwarf absorption features, and to show the structure in the HeI $\lambda 5876$ line; the orbital motion of the M dwarf is clearly seen. The scaling of the 
lower panel brings out the complex structure in the H$\alpha$ emission.


\section{Distance}
\label{sec:dist}

We can estimate the distance to \hs\ using the secondary star's contribution to the spectrum and our knowledge of the orbital period \porb. For a secondary 
star of mass \msec\, at a fixed \porb, the Roche lobe radius $R_2$ is proportional to $\msec^{1/3}$, and is almost independent of the primary mass \mwd\, 
\citep{beuermannetal98-1}. We do not know \msec, but we can estimate it using evolutionary models tabulated by \citet{baraffe+kolb00-1}; these suggest that 
the secondary is between 0.23 and 0.56 \msun. At this \porb, Eq.\,1 of \citet{beuermannetal98-1} then implies $R_2\,=\,0.47\,\pm\,0.07\,\rsun$. 
\citet{beuermannetal99-1} tabulate absolute magnitudes and radii for late type dwarfs as a function of spectral class, which implies a surface brightness for 
each star. In the range of spectral type we see here, these correspond to $M_V\,=\,8.8\,\pm\,0.7$ for a $1$\rsun\, star, where the uncertainty includes both 
the spectral type uncertainty and the scatter among the tabulated points. Combining this with the radius yields an estimate of $M_V = 10.4 \pm 0.8$ for the 
secondary. The Galactic coordinates of \hs\ are $(l,b)\,=\,(120^{\circ},20^{\circ})$; at this location, \citet{schlegeletal98-1} estimate $E(B-V)\,=\,0.19$ to 
the edge of the Galaxy. Assuming that our object lies outside most of the dust, and taking the M-dwarf contribution to the spectrum as $V\,=\,19.0\,\pm\,0.4,$ 
then yields an extinction-corrected distance modulus of $(m - M)_0\,=\,8.0\,\pm\,0.9$, corresponding to a distance of $400 (+200, -140)$ pc. Note carefully 
that this estimate makes no assumption that the secondary follows a main-sequence mass-radius relation; it assumes only that the secondary's spectral type is 
a reliable guide to its surface brightness, and that it fills its Roche lobe.


\section{Outburst behaviour}
\label{sec:outbeh}

We intensively monitored \hs\ for about 50 days, starting from the 1st of April 2009 using small aperture telescopes. Four outbursts have been recorded during 
this period, indicating a recurrence time of $\sim\,12-14\,\mathrm{d}$. Prominent in Fig.\,\ref{fig:outburst} is a ``long'' outburst, lasting 
$\sim\,11-12\,\mathrm{d}$, followed by a seemingly ``short'' outburst. This could be a hint towards a bimodal distribution of the outburst duration, observed 
in many dwarf novae \citep[see e.g.][]{szkody+mattei84-1,aketal02-1}. Further observations are required to establish a more accurate recurrence time and to 
check the consistency of the ``long'' and ``short'' outburst succession.

We have inspected v7.12 (2009) of the Ritter and Kolb catalogue \citep{ritter+kolb03-1} and compiled a list of U\,Gem-type dwarf novae (UG) and Z\,Cam-type 
stars (ZC) that are found in the range $4\,\mathrm{h}\,<\,\porb\,<\,5\,\mathrm{h}$. Only systems with confirmed UG/ZC status and with a quoted outburst 
recurrence period were considered. This left us with a list of 22 systems (out of the 39 listed in R\&K in this \porb\, range). In this list ZC systems 
dominate the short end of the outburst recurrence period distribution ($11-18\,\mathrm{d}$), while UG systems tend to have longer intervals between outbursts 
($16-150\,\mathrm{d}$). Our inferred outburst recurrence period places \hs\ in the ZC region. However, as there has been no recorded standstill (the hallmark 
of ZC systems), its identification as either a UG or a ZC remains ambiguous.


\section{Estimates of the binary properties}
\label{sec:binprop}

The standard treatment of eclipsing CVs \citep[see e.g.][]{woodetal86-1,woodetal92-1,littlefairetal06-1} involves the identification of the contact points of 
the white dwarf, the bright spot and the accretion disc. The corresponding phase-widths are then used to place firm (geometrical) constraints on the mass ratio 
$q$ and the inclination angle $i$ \citep[e.g.][]{bailey79-1,horne85-1} and deduce information about the extend and location of the bright spot and the size of 
the accretion disc. Flickering can hinder attempts to identify the contact points. Averaging together many light curves is an often applied solution 
\citep[see e.g.][for the case of IP\,Peg]{copperwheatetal10-1}.

Although breaks in slope are seen in the light curve of \hs, the available data set is not of sufficient quality and time-resolution to unambiguously identify 
the different contact points. Hence, the exact eclipse geometry of \hs\, remains unclear.

In an attempt to constrain the parameter space (albeit roughly) we have to rely on theoretical predictions and empirical evidence from the observed CV 
population, coupled with the limited information that can be extracted from the light curves.

Following the procedure outlined in detail in \citet{dhillonetal91-1} (D91) the radius of the accretion disc can be determined as a function of the binary 
separation, $q$ and $i$, for a given eclipse half-width at maximum intensity $\Delta\phi$ (essentially timing the first and last contacts of eclipse and 
dividing by two). $\Delta\phi$ was determined by eye to be $\Delta\phi\,=\,0.1\,\pm\,0.02$. The large error is due to the fact that the exact beginning and 
end of the eclipse are uncertain because of flickering.

\begin{figure}
\plottwo{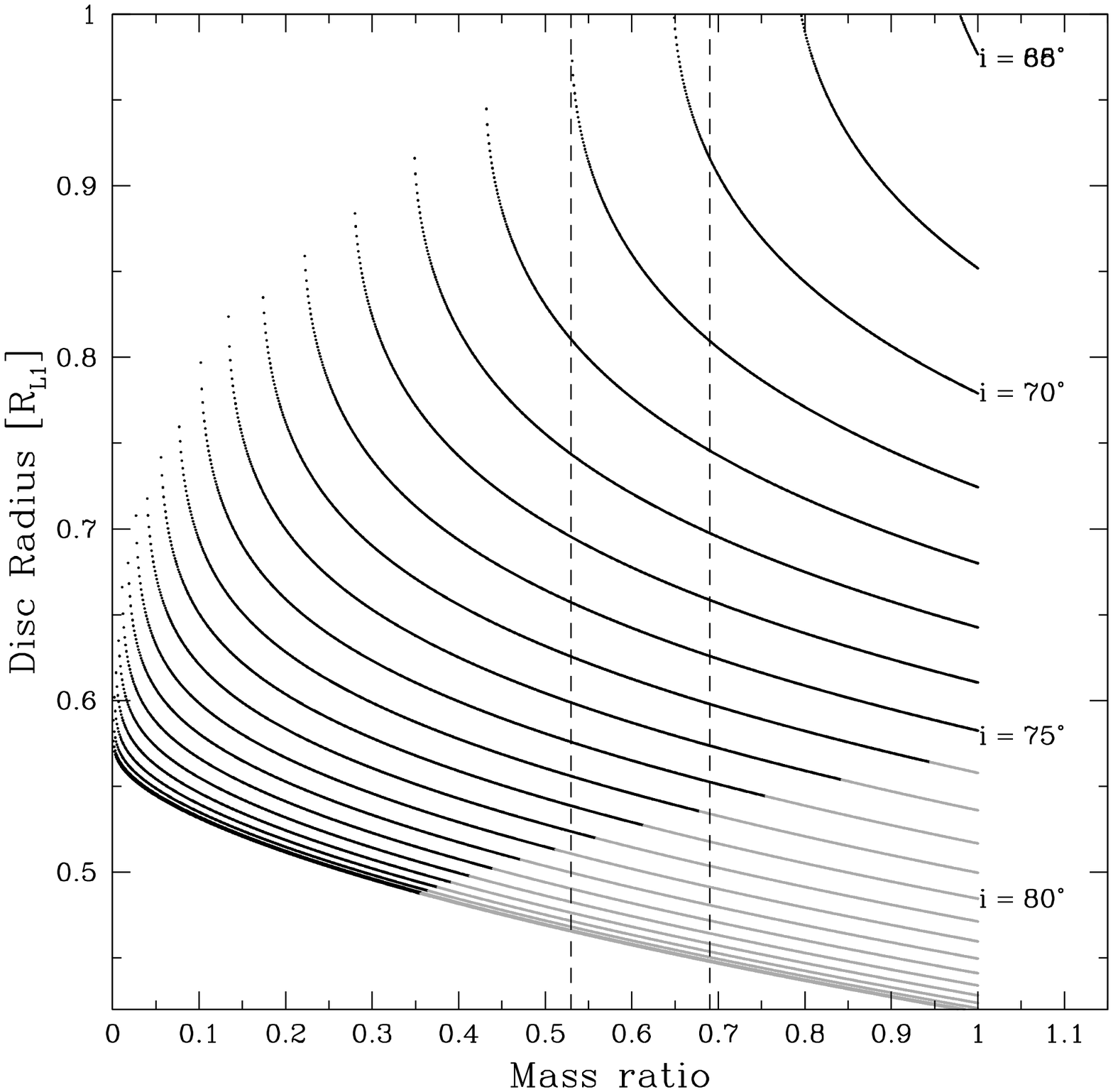}{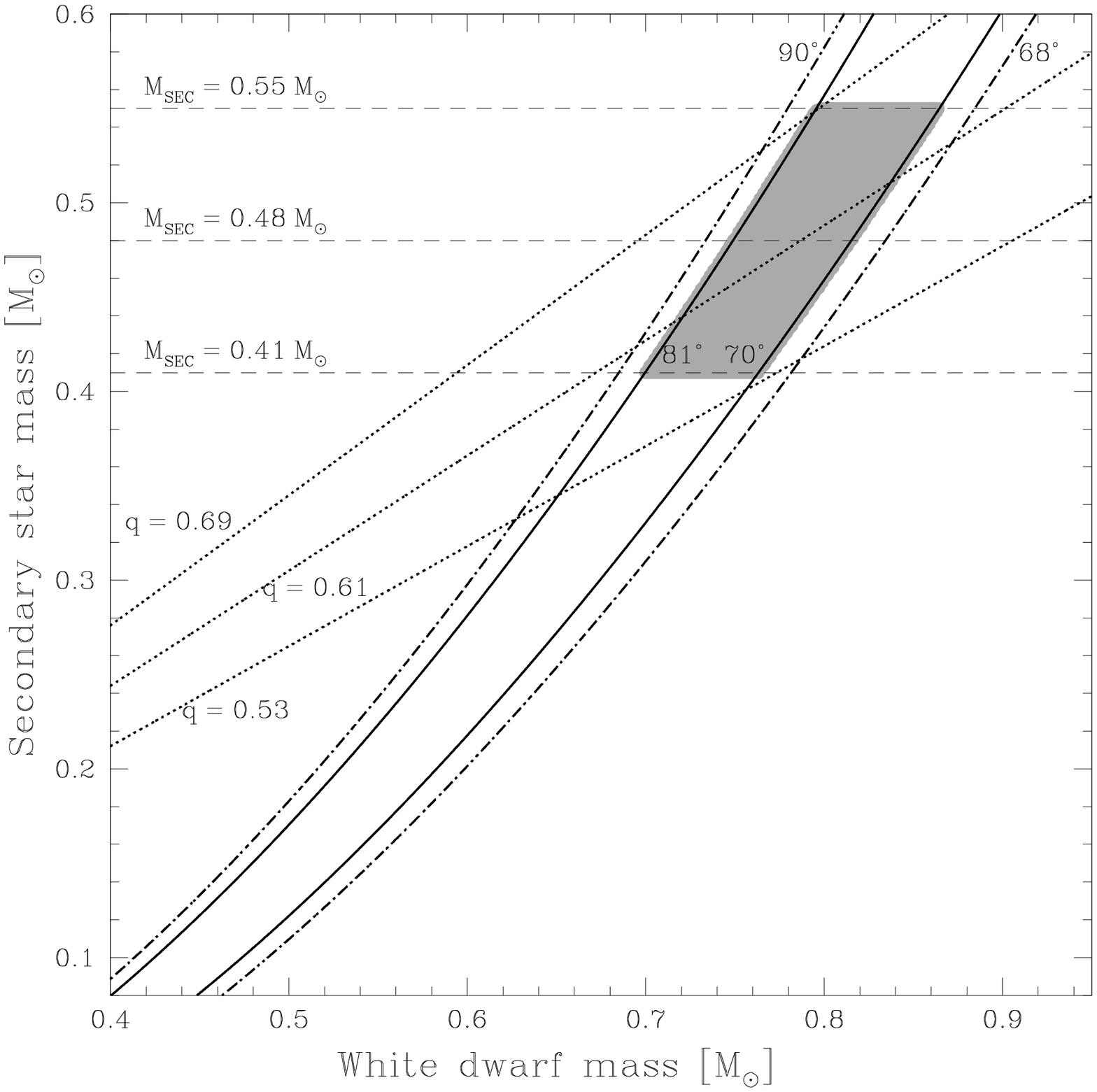}
\caption{Left panel: Accretion disc radius, in units of [$R_{L_{1}}$], as a function of $q$ and $i$; the curves are bound below - grey coloured - by the 
requirement of a partial disc eclipse, while the vertical dashed lines correspond to a spectroscopic constraint on $q$ (see text for details). Right panel: 
constraints on the masses of the binary components; Eq.\,\ref{eq:mf} plotted for $i\,=\,68^{\circ},90^{\circ}$ (dash-dot curves) and 
$i\,=\,70^{\circ},81^{\circ}$ (solid curves), constraints on the secondary mass assuming the mass-period relation of \citet{smith+dhillon98-1} (dashed lines) 
and constraints on the mass ratio $q$ assuming $K_{\mathrm{em}}\,=\,\kwd$ (dotted lines). Shaded in grey is the allowed parameter space, under the previous 
assumptions (see text for details).\label{fig:binparam}}
\end{figure}

The left panel of Fig.\,\ref{fig:binparam} shows the disc radius $R_{\mathrm{D}}$ (in units of the distance between the primary and the inner Lagrangian point, $R_{L_{1}}$) 
calculated using Equations 3, 4 and 5 of D91, for $0\,\le\,q\,\le\,1$ and various inclination angles. The curves are bound above by the requirement that 
$R_{\mathrm{D}}\,\le\,R_{L_{1}}$ and below by the requirement for a partial disc eclipse, satisfied if the disc radius is larger than the half-cord of the 
secondary (shown by the change in line colour in the left panel of Fig.\,\ref{fig:binparam}). This allows us to place a strict lower limit for the inclination 
angle to be $i_{\mathrm{min}}\,=\,68^{\circ}$. However, the upper limit of $i$ and the possible values of $q$ remain unconstrained.

Using the mass function,

\begin{equation}
\label{eq:mf}
 f\left(\mwd\right)=\frac{\left(\mwd\sin i\right)^{3}}{\left(\mwd+\msec\right)^{2}}=\frac{P_{\mathrm{orb}}K^{3}_\mathrm{sec}}{2\pi G}\quad \leq\,\mwd
\end{equation}

\noindent
we can transform a given $\left(q,i\right)$ pair to a unique $\left(\mwd,\msec\right)$ pair. The right panel of Figure\,\ref{fig:binparam} shows 
Eq.\,\ref{eq:mf} calculated for $i_{\mathrm{min}}\,=\,68^{\circ}$ and $i_{\mathrm{max}}\,=\,90^{\circ}$, over a wide range in secondary mass, 
$0.1\,\le\,\msec\,[\msun]\,\le\,0.6$. Allowed $\left(\mwd,\msec\right)$ pairs are located between the two dash-dot curves.

We can further narrow down the parameter space, by making two assumptions:

\begin{enumerate}
 \item The secondary follows the mass-period relation of \citet{smith+dhillon98-1}; their Equation\,8 (power-law fit) yields 
$\msec\,=\,0.43\,\pm\,0.07\,\msun$, while their Equation\,9 (linear fit) yields $\msec\,=\,0.48\,\pm\,0.07\,\msun$ (Fig.\,\ref{fig:binparam}, 
right panel, dashed horizontal lines). An average of these values is in perfect agreement with the value of $\msec\,=\,0.45\,\msun$ predicted by the revised 
model track of \citet{kniggeetal11-1} for this orbital period.
 \item The radial velocity variation of the emission lines tracks the motion of the white dwarf,
       so $K_{\mathrm{em}}\,=\,\kwd\,=\,145\,\pm\,9\,\kms$ and, therefore, 
       $q\,=\,\kwd/\ksec\,=\,0.61\,\pm\,0.08$ (Fig.\,\ref{fig:binparam}, right panel, dotted lines).
\end{enumerate}

While the latter is a frequently adopted assumption in CV research, it has to be viewed with a certain amount of caution, see e.g. \citet{shafter83-1} and 
\citet{thorstensen00-1}. An encouraging fact in the case of \hs\ is that the phasing of the emission lines is consistent with the eclipse ephemeris. The 
constraint on $q$ imposes a narrower range of inclination angles $70^{\circ}\,\le\,i\,\le\,81^{\circ}$ (Fig.\,\ref{fig:binparam}, left panel, dashed vertical 
lines). If these assumptions are indeed correct, then the allowed $\left(\mwd,\msec\right)$ pairs are indicated as the grey shaded area 
in the right panel of Fig.\,\ref{fig:binparam}.


\section{Discussion and Conclusions}
\label{sec:concl}

In this paper, we identified HS\,2325+8205 as an eclipsing, frequently outbursting dwarf nova above the CV orbital period gap and presented our photometric and 
spectroscopic data. We used these data to measure the orbital period and the radial velocity of the secondary star, as well as provide initial estimates on the
binary parameters. With the photometric data at hand it remains unclear whether the white dwarf is fully eclipsed or not.  The shallow eclipse depth 
could suggest that it is not eclipsed at all.  High time resolution and signal-to-noise data at quiescence are needed in order to unambigously identify 
the eclipse geometry. 

The short outburst cycle of $\sim\,12-14\,\mathrm{d}$ makes \hs\ a plausible Z\,Cam-type candidate. If confirmed, it will be only the third known eclipsing 
Z\,Cam system, after EM\,Cyg \citep[e.g.][and references therein]{northetal00-1} and AY\,Psc \citep[e.g.][and references therein]{guelsecenetal09-1}.

\hs, with its high declination (circumpolar, ideal for observers in the northern hemisphere), short outburst recurrence period and magnitude range (accessible 
with 2-4m class telescopes) offers an excellent target for systematic follow-up observations. Simultaneous high-time resolution photometric and spectroscopic 
observations can provide unique insight in the changes in the structure of the accretion disc, through techniques such as eclipse mapping \citep{horne85-1} and 
Doppler tomography \citep{marsh+horne88-1}.

Furthermore, the outbursts of \hs\, can be picked-up with relative ease by small aperture telescopes enabling the accumulation of a very long baseline of 
outburst data, which can then be compared to the predictions of the disc instability model. We strongly encourage observers from around the world to frequently 
monitor the system and put the Z\,Cam-type scenario to the test.


\acknowledgments

JRT gratefully acknowledges support from the U.S. National Science Foundation, through grants AST-0307413 and AST-0708810. Based in part on observations made with the Nordic Optical Telescope, 
operated on the island of La Palma jointly by Denmark, Finland, Iceland, Norway, and Sweden, in the Spanish Observatorio del Roque de los Muchachos of the Instituto de Astrof\'isica de Canarias; 
on observations collected at the Centro Astron\'omico Hispano Alem\'an (CAHA) at Calar Alto, operated jointly by the Max-Planck Institut f\"ur Astronomie and the Instituto de Astrof\'isica de 
Andaluc\'ia (CSIC); on observations made at the 1.2m telescope, located at Kryoneri Korinthias, and owned by the National Observatory of Athens, Greece; and on observations obtained at the MDM 
Observatory, operated by Dartmouth College, Columbia University, Ohio State University, and the University of Michigan.




\clearpage

\begin{deluxetable}{cccccccccc}
\tabletypesize{\scriptsize}
\setlength{\tabcolsep}{0.75ex}
\tablecaption{Log of the observations with large aperture ($>\,1\rm{m}$) telescopes.  Given is the run ID, the date of observation, the start and end of 
observation in HJD, the telescope and the filter (photometry) or grating (spectroscopy) used. ``Frames'' denotes the number of frames collected, ``Ecl'' the 
number of eclipses observed, ``Mag'' the out-of-eclipse magnitude and ``Depth'' the eclipse depth in magnitudes. Information about the telescopes and 
instruments used is provided at the bottom of the Table.\label{tab:jobs}}
\tablewidth{0pt}
\tablehead{
\colhead{ID} & \colhead{Date} & \colhead{HJD range$^{\rm{a}}$} & \colhead{Telescope} & \colhead{Filter/Grating} & \colhead{Exp. time} & \colhead{Frames} & 
\colhead{Ecl.} & \colhead{Mag.} & \colhead{Depth} \\ & & & & & [s] & & & & [mag]} 
\startdata
01 & 2003 Sep 05 & 2888.255 - 2888.518 & KY$^{\rm{b}}$ & clear & 30 & 646 & 1 & 14.9 & 0.4 \\
02 & 2004 Jun 10 & 3167.429 - 3167.467 & KY & clear & 30 &  90 & 0 & 15.1 &   - \\
03 & 2004 Jun 11 & 3168.413 - 3168.593 & KY & clear & 30 & 390 & 1 & 15.7 & 0.8 \\
04 & 2004 Jun 12 & 3169.451 - 3169.569 & KY & clear & 30 & 270 & 0 & 16.1 &   - \\
05 & 2004 Jul 25 & 3212.455 - 3212.607 & KY & clear & 30 & 389 & 1 & 16.0 & 0.8 \\
06 & 2004 Jul 27 & 3214.298 - 3214.573 & KY & clear & 30 & 693 & 2 & 16.2 & 0.8 \\
07 & 2004 Oct 21 & 3300.212 - 3300.493 & KY & clear & 20 & 997 & 2 & 16.4 & 0.8 \\
08 & 2004 Oct 22 & 3301.213 - 3301.478 & KY & clear & 20 & 897 & 1 & 15.9 & 0.7 \\
09 & 2004 Oct 23 & 3302.214 - 3302.457 & KY & clear & 20 & 858 & 1 & 14.6 & 0.4 \\
10 & 2005 Sep 05 & 3619.333 - 3619.546 & CA22$^{\rm{c}}$ & clear & 10 & 572 & 1 & 14.4 & 0.4 \\
11 & 2005 Sep 11 & 3624.727 - 3624.804 & HT$^{\rm{d}}$ & 600 l/mm & 360/480 & 17 & - & - & - \\ 
12 & 2005 Sep 12 & 3625.649 - 3625.909 & HT & 600 l/mm & 360/480 & 31 & - & - & - \\
13 & 2005 Sep 15 & 3629.599 - 3629.668 & NOT$^{\rm{e}}$ & clear & 4 & 564 & 1 & 17 & 0.7 \\
14 & 2005 Sep 16 & 3630.550 - 3630.715 & NOT & clear & 4 & 1308 & 1 & 16.5 & 0.6 \\
15 & 2006 Aug 23 & 3971.321 - 3971.621 & KY & clear & 30 & 727 & 1 & 14.5 & 0.4 \\
16 & 2006 Oct 28 & 4037.248 - 4037.504 & KY & clear & 30 & 595 & 1 & 16.3 & 0.8 \\
17 & 2007 Jan 24 & 4124.604 - 4124.610 & HT & 600 l/mm & 360/480 & 2 & - & - & - \\
\enddata
\tablenotetext{a}{Start to end of observation, in the format HJD\,-\,2450000.}
\tablenotetext{b}{1.2m Kryoneri telescope (KY), CCD SI-502, 516\,x\,516, FOV $2.5'$\,x\,$2.5'$}
\tablenotetext{c}{2.2m Calar Alto telescope (CA22), CAFOS, CCD SITe 2k\,x\,2k, FOV $16'$\,x$\,16'$}
\tablenotetext{d}{2.4m Hiltner Telescope (HT), CCD SiTe 2k\,x\,2k}
\tablenotetext{e}{2.5m Nordic Optical Telescope (NOT), ALFOSC, CCD EEV42-40 2k\,x\,2k, FOV $6.5'$\,x\,$6.5'$}
\end{deluxetable}


\clearpage

\begin{deluxetable}{cccccccccccc}
\tabletypesize{\scriptsize}
\setlength{\tabcolsep}{0.75ex}
\tablecaption{Mid-eclipse times and their errors, the difference between observed and computed eclipse times and the cycle number using the ephemeris provided 
in Eq.\,\ref{eq:ephe}. Times are calculated on a UTC timescale and given in the format $\mathrm{HJD}\,-\,2450000$. \label{tab:mideclt}}
\tablewidth{0pt}
\tablehead{
\colhead{$T_{0}$} & \colhead{Error} & \colhead{O-C} & \colhead{Cycle} & \colhead{$T_{0}$} & \colhead{Error} & \colhead{O-C} & \colhead{Cycle} 
& \colhead{$T_{0}$} & \colhead{Error} & \colhead{O-C} & \colhead{Cycle} \\ \colhead{[HJD]} & \colhead{[HJD]} & \colhead{[s]} & & \colhead{[HJD]} & 
\colhead{[HJD]} & \colhead{[s]} & & \colhead{[HJD]} & \colhead{[HJD]} & \colhead{[s]} & } 
\startdata
2888.425480 &  0.000405 &  -6 &     0 &  4320.476770 &  0.000463 &   3 &  7369 &  4940.598350 &  0.000463 &  10 &  10560 \\
3168.461560 &  0.000405 &  -4 &  1441 &  4335.440630 &  0.000463 &  12 &  7446 &  4941.375690 &  0.000775 &  10 &  10564 \\
3212.575440 &  0.000405 &  -9 &  1668 &  4358.371920 &  0.000463 &  -4 &  7564 &  4941.569840 &  0.000463 &  -6 &  10565 \\
3214.324580 &  0.000405 &   2 &  1677 &  4379.360110 &  0.000463 &   1 &  7672 &  4942.347170 &  0.001157 &  -7 &  10569 \\
3214.518890 &  0.000405 &  -0 &  1678 &  4393.352210 &  0.000463 &   2 &  7744 &  4942.541460 &  0.000463 & -11 &  10570 \\
3300.220510 &  0.000289 &   7 &  2119 &  4451.458200 &  0.000810 &  -1 &  8043 &  4943.513250 &  0.001157 &  -0 &  10575 \\
3300.414840 &  0.000289 &   7 &  2120 &  4510.535910 &  0.000463 &   0 &  8347 &  4944.485020 &  0.001157 &   8 &  10580 \\
3301.386470 &  0.000289 &   3 &  2125 &  4524.527990 &  0.000810 &  -1 &  8419 &  4945.456530 &  0.001157 &  -6 &  10585 \\
3302.357990 &  0.000289 & -10 &  2130 &  4544.544400 &  0.000463 &  -5 &  8522 &  4945.650900 &  0.000660 &  -3 &  10586 \\
3619.512060 &  0.000231 &  -1 &  3762 &  4560.479940 &  0.000810 &   5 &  8604 &  4945.845190 &  0.000660 &  -7 &  10587 \\
3629.617580 &  0.000116 &  10 &  3814 &  4927.383490 &  0.000775 &   0 &  1049 &  4946.428210 &  0.001157 &  -6 &  10590 \\
3630.589230 &  0.000116 &   8 &  3819 &  4933.602120 &  0.000810 &  -6 &  1052 &  4947.399870 &  0.000775 &  -7 &  10595 \\
3971.452040 &  0.000405 &  11 &  5573 &  4934.573920 &  0.000810 &   4 &  1052 &  4948.371720 &  0.000660 &   8 &  10600 \\
4037.331340 &  0.000405 &   2 &  5912 &  4935.351320 &  0.000775 &  10 &  1053 &  4963.529620 &  0.000810 &  -8 &  10678 \\
4289.577510 &  0.000463 &  -3 &  7210 &  4935.545640 &  0.001007 &   8 &  1053 &  4964.695730 &  0.000810 &   0 &  10684 \\
4312.508950 &  0.000463 &  -6 &  7328 &              &           &     &       &              &           &     &        \\
\enddata
\end{deluxetable}


\clearpage

\clearpage

\clearpage

\clearpage

\clearpage

\clearpage

\end{document}